\documentclass[prb,showpacs,twocolumn]{revtex4}

\usepackage{amsmath,amssymb,epsfig,epsf}

\usepackage{color}

\begin{document}

\title
{Doping Dependence of Correlation Effects in K$_{1-x}$Fe$_{2-y}$Se$_2$
Superconductor: LDA$^\prime$+DMFT Investigation.}

\author {I.A. Nekrasov$^1$, N.S. Pavlov$^1$, M.V. Sadovskii${^{1,2}}$}

\affiliation
{$^1$Institute for Electrophysics, Russian Academy of Sciences, Ural Branch,\\ Amundsen str. 106, Ekaterinburg, 620016, Russia\\
$^2$Institute for Metal Physics, Russian Academy of Sciences, Ural Branch, \\  S. Kovalevskaya str. 18, Ekaterinburg, 620990, Russia}

\begin{abstract}

We present detailed LDA$^\prime$+DMFT investigation of doping dependence of correlation effects in novel
K$_{1-x}$Fe$_{2-y}$Se$_2$ superconductor. Calculations were performed at four different hole doping levels,
starting from hypothetical stoichiometric composition with total number of electrons equal to 29 per unit cell
through 28 and 27.2 electrons towards the case of 26.52,
which corresponds to chemical composition K$_{0.76}$Fe$_{1.72}$Se$_2$ studied in recent ARPES experiments.
In general case the increase of hole doping leads to quasiparticle bands in wide energy window $\pm$2 eV
around the Fermi level  becoming more broadened by lifetime effects, while correlation induced compression of
Fe-3d LDA$^\prime$ bandwidths stays almost the same and of the order of $\sim$1.3 for all hole concentrations.
However close to the Fermi level situation is more complicated.
Here in the energy interval from -1.0 eV to 0.4 eV the bare Fe-3d LDA$^\prime$ bands are compressed by
significantly larger renormalization factors up to 5 with hole doping increase, while the value of Coulomb interaction remains the same.
This fact manifests the increase of correlation effects with hole doping in K$_{1-x}$Fe$_{2-y}$Se$_2$ system.
Moreover in contrast to typical pnictides K$_{1-x}$Fe$_{2-y}$Se$_2$ does not have well defined quasiparticle bands
on the Fermi leves but pseudogap like ``dark'' region instead. 
We also find that with the growth of hole doping Fe-3d orbitals of various symmetries
are affected by correlations in a different way in different parts of Brillouin zone.
To illustrate this we determine quasiparticle mass renormalization factors and energy shifts,
which transform the bare Fe-3d LDA$^\prime$ bands of various symmetries into
LDA$^\prime$+DMFT quasiparticle bands. These renormalization factors effectively mimic more complicated
energy dependent self-energy effects and can be used to analyze the available ARPES data.

\end{abstract}

\pacs{71.20.-b, 71.27.+a, 71.28.+d, 74.20.Fg, 74.25.Jb,  74.70.-b}

\maketitle

\section{Introduction}

The discovery of iron based high-temperature superconductors [\onlinecite{kamihara_08}]
high-temperature superconductors stimulated quite intensive research work
[\onlinecite{UFN_90,Hoso_09,FeSe,Mazin,Kord}].
Recently another class of high-T$_c$ superconductors  isostructural to 122-family of iron pnictides
was discovered -- iron chalcogenides K$_x$Fe$_2$Se$_2$ [\onlinecite{Guo10}],
Cs$_x$Fe$_2$Se$_2$ [\onlinecite{Krzton10}] and (Tl,K)Fe$_x$Se$_2$ [\onlinecite{Fang}].
Values of superconducting critical temperatures T$_c$ are comparable for both
families of pnictides and chalcogenides and are about 30-50K [\onlinecite{rott,ChenLi,Chu,Bud}].
Further interest to these chalcogenides was stimulated by
experimental observation of rather nontrivial antiferromagnetic
ordering with very high Neel temperature of about 550K and Fe vacancies ordering
in the same range of temperatures in K$_{0.8}$Fe$_{1.6}$Se$_2$
(the so called 245 phase) [\onlinecite{Fe_order}].
Despite intensive experimental work there is still no consensus on the composition of the phase,
responsible for high-$T_c$ superconductivity in these systems.
The most common point of view is that KFe$_2$Se$_2$ (122 phase) is the
parent compound for superconductivity, while 245 phase is insulating
[\onlinecite{MISM,Kord,Wen}]. Some other phases in this system were also reported [\onlinecite{WLi}].
Below we shall concentrate on electronic structure calculations for the parent 122 phase with
different levels of hole doping.

Crystallographically pnictides AFe$_2$As$_2$, and chalcogenides Fe(Se,Te), AFe$_2$Se$_2$
are quite similar with main structural motiff -- layers of Fe(As,Se)$_4$ tetrahedra.
Recently reported theoretical electronic band structures of
Fe(Se,Te) [\onlinecite{SinghFeSe}] and AFe$_2$As$_2$ [\onlinecite{Nekr2,Shein,Krell}]
are found to be nearly identical to each other especially for those bands crossing Fermi level.
AFe$_2$As$_2$ and AFe$_2$Se$_2$ compounds are simply isostructural.
However LDA (local density approximation) electronic structure of
AFe$_2$Se$_2$ differs quite remarkably from that of AFe$_2$As$_2$ as was directly shown in
Refs.~[\onlinecite{Shein_kfese,Nekr_kfese,MISM}].

From the very beginning of studies of iron based superconductors it was recognized,
that the account of electronic correlations on Fe sites is rather essential for
correct description of the physics of pnictide materials
[\onlinecite{Haule,Craco,Shorikov,Ba122_DMFT}].
To this end LDA+DMFT hybrid computational scheme [\onlinecite{LDADMFT}]
was employed. The main conclusion was that correlations lead
to simple narrowing (compression) of LDA bandwidth
by the factor of the order of 2 or 3. This observation agrees rather well
with variety of angular resolved photoemission spectroscopy (ARPES)
data on AFe$_2$As$_2$ compounds \onlinecite{Kord}. Fermi surface maps obtained from
ARPES experiments for AFe$_2$As$_2$ are quite similar to those obtained from simple LDA
calculations: there are 2 or 3 hole cylinders around $\Gamma$-pointin the Brillouin zone
and 2 electron Fermi surface sheets around ($\pi,\pi$) point.

Up to now there are only few LDA+DMFT papers devoted to Fe chalcogenides [\onlinecite{DMFT_FeSe,DMFT_AFeSe}].
Recently we performed the investigation of electronic structure of hole doped iron chalcogenide
K$_{0.76}$Fe$_{1.72}$Se$_2$
in the normal phase [\onlinecite{kfese_dmft}], inspired by available ARPES data,
for this system [\onlinecite{ARPES_AFeSe}], especially those obtained in Ref.~[\onlinecite{exp}],
using both the standard LDA+DMFT and the novel LDA$^\prime$+DMFT computational approach [\onlinecite{CLDA,CLDA_long}].
The results of our calculations agree rather well with the general picture of ARPES, obtained in Ref. [\onlinecite{exp}],
with LDA$^\prime$+DMFT showing slightly better agreement. We showed that this iron chalcogenide is actually more strongly
correlated in the sense of bandwidth renormalization (energy scale compression by the factor of about 5 in the energy
interval $\pm$1.5~eV) around the Fermi level), than the typical iron pnictides (with compression factor of about 2 or 3
[\onlinecite{Kord}]), though the Coulomb interaction strength is almost the same in both families.
Moreover the K$_{0.76}$Fe$_{1.72}$Se$_2$ system demonstrates absence of well defined quasiparticle
bands on the Fermi level in contract to pnictides.

In this paper we continue our LDA$^\prime$+DMFT study of  system.
We investigate the evolution of correlation effects upon hole doping, performing LDA$^\prime$ calculations at four
doping levels:
hypothetical stoichiometric composition with 29 valence electrons per unit cell
through intermediate values of valence electrons 28 and 27.2 downto experimentally obtained
composition K$_{0.76}$Fe$_{1.72}$Se$_2$ with 26.52 electrons per unit cell.
We shall demonstrate that the actual doping dependence of correlation effects on the electronic structure in
iron chalcogenides is apparently more complicated than in iron pnictides and does not reduce to the simple
picture of universal bandwidth renormalization (compression).

The paper is organized in a following manner.
In Sec.~\ref{cd} we discuss crystallographic structure, methodological and computational details
of LDA$^\prime$+DMFT. Comparative study of LDA$^\prime$ bands and LDA$^\prime$+DMFT spectral function maps
within wide and narrow energy intervals around the Fermi level, together with orbitally resolved densities
of states are presented in Sec.~\ref{rad}. Finally we summarise our results in Sec.~\ref{concl}.
\begin{figure}[ht]
\includegraphics[width=.5\textwidth]{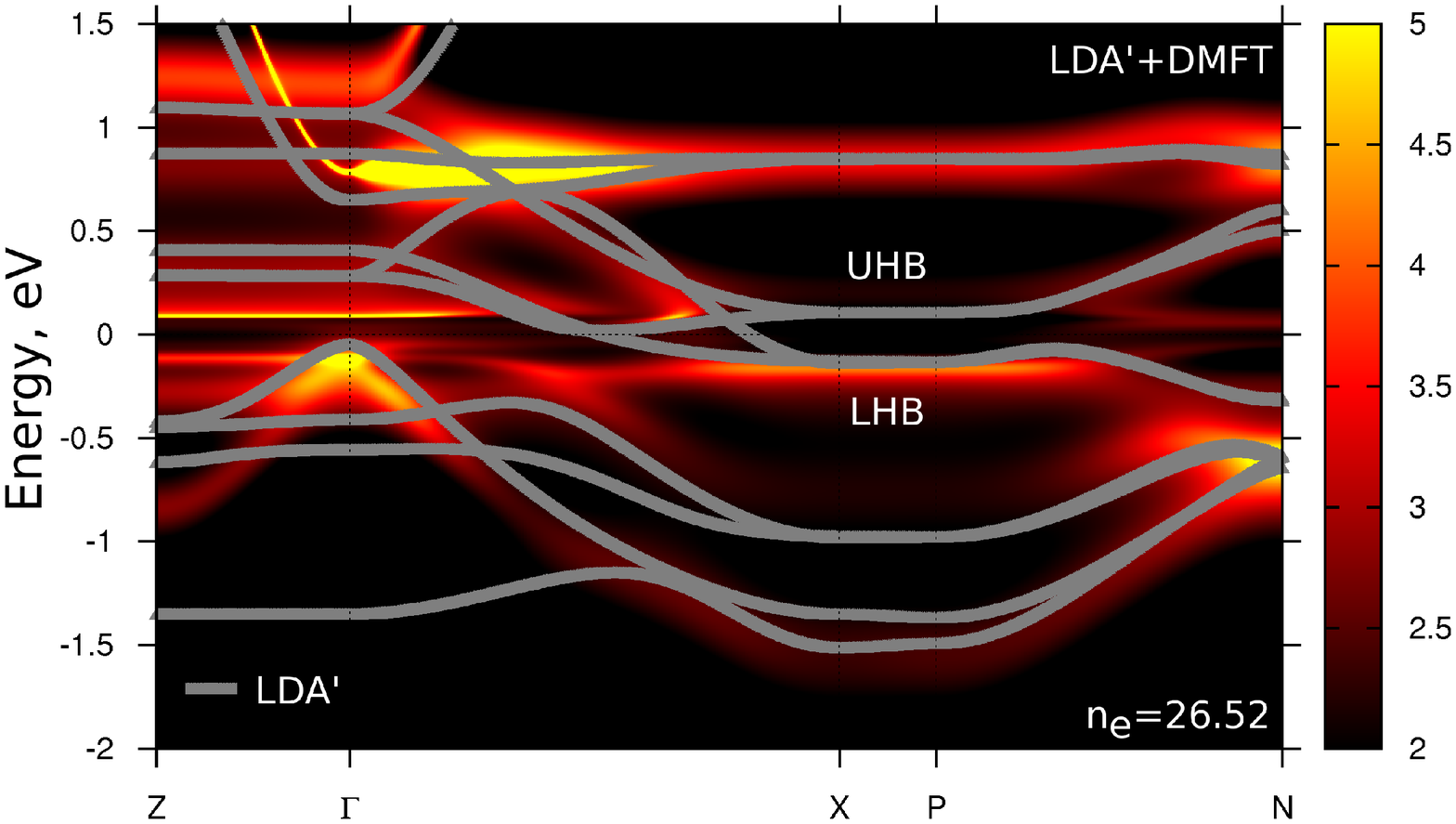}
\includegraphics[width=.5\textwidth]{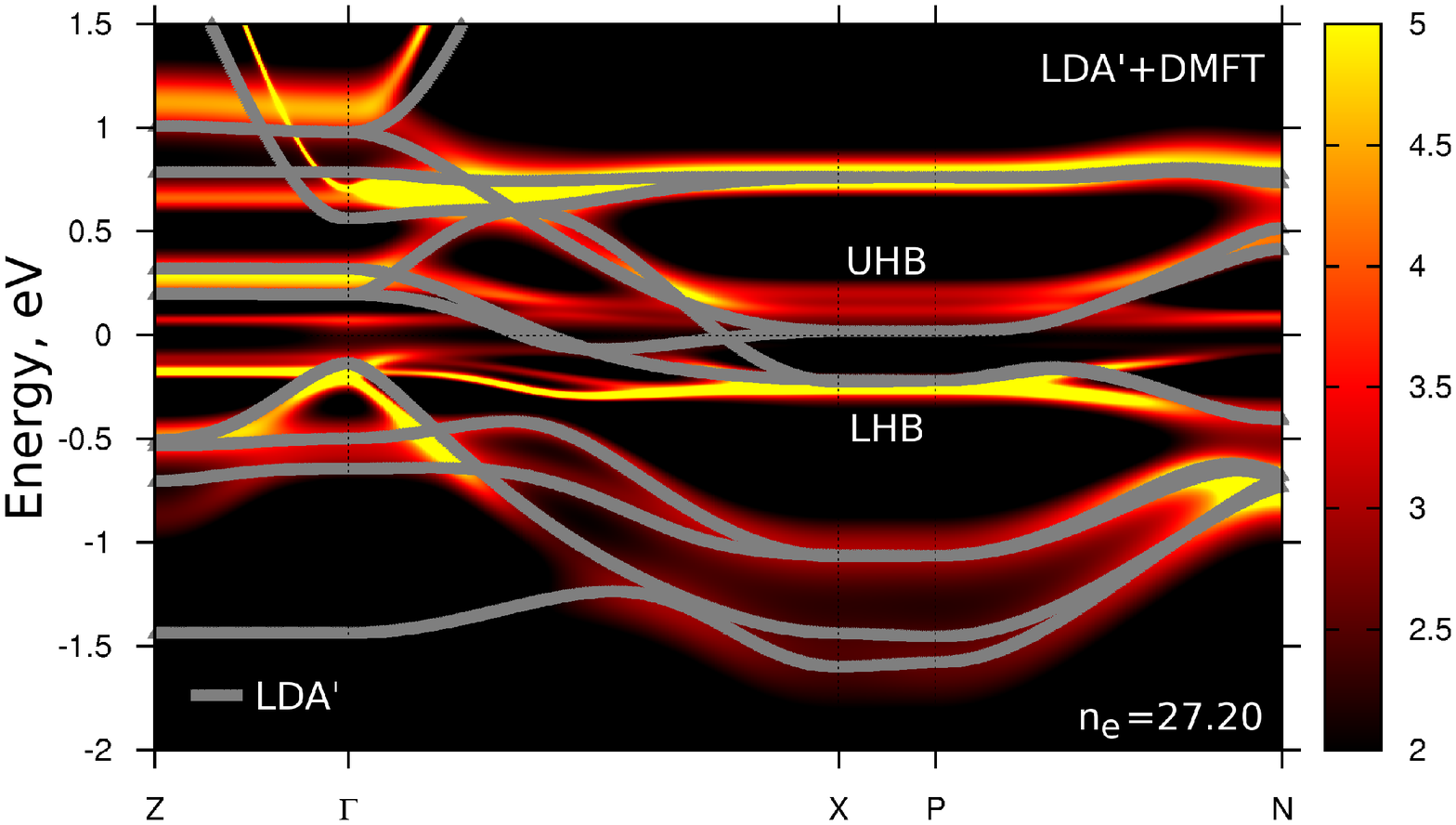}
\includegraphics[width=.5\textwidth]{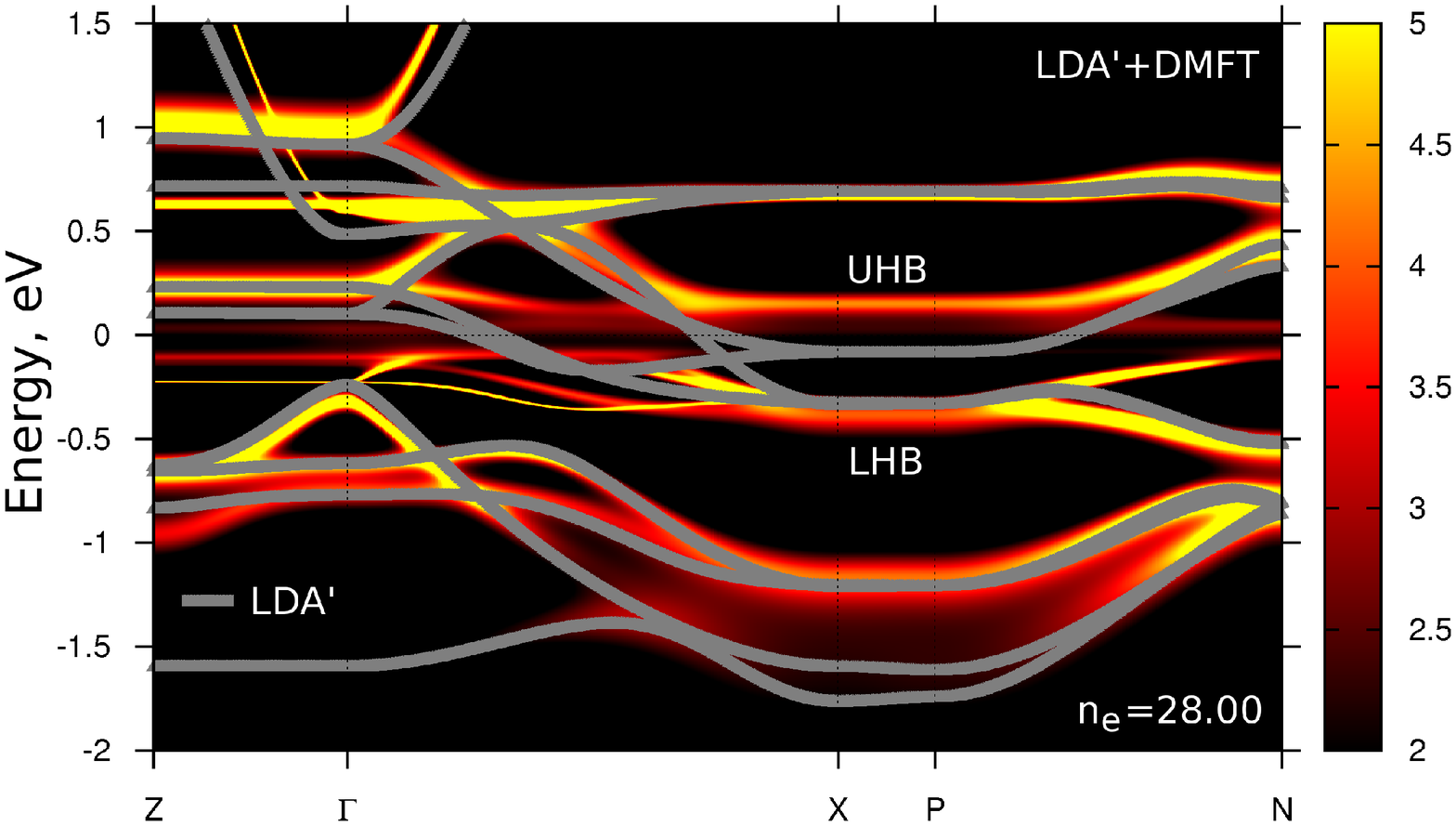}
\includegraphics[width=.5\textwidth]{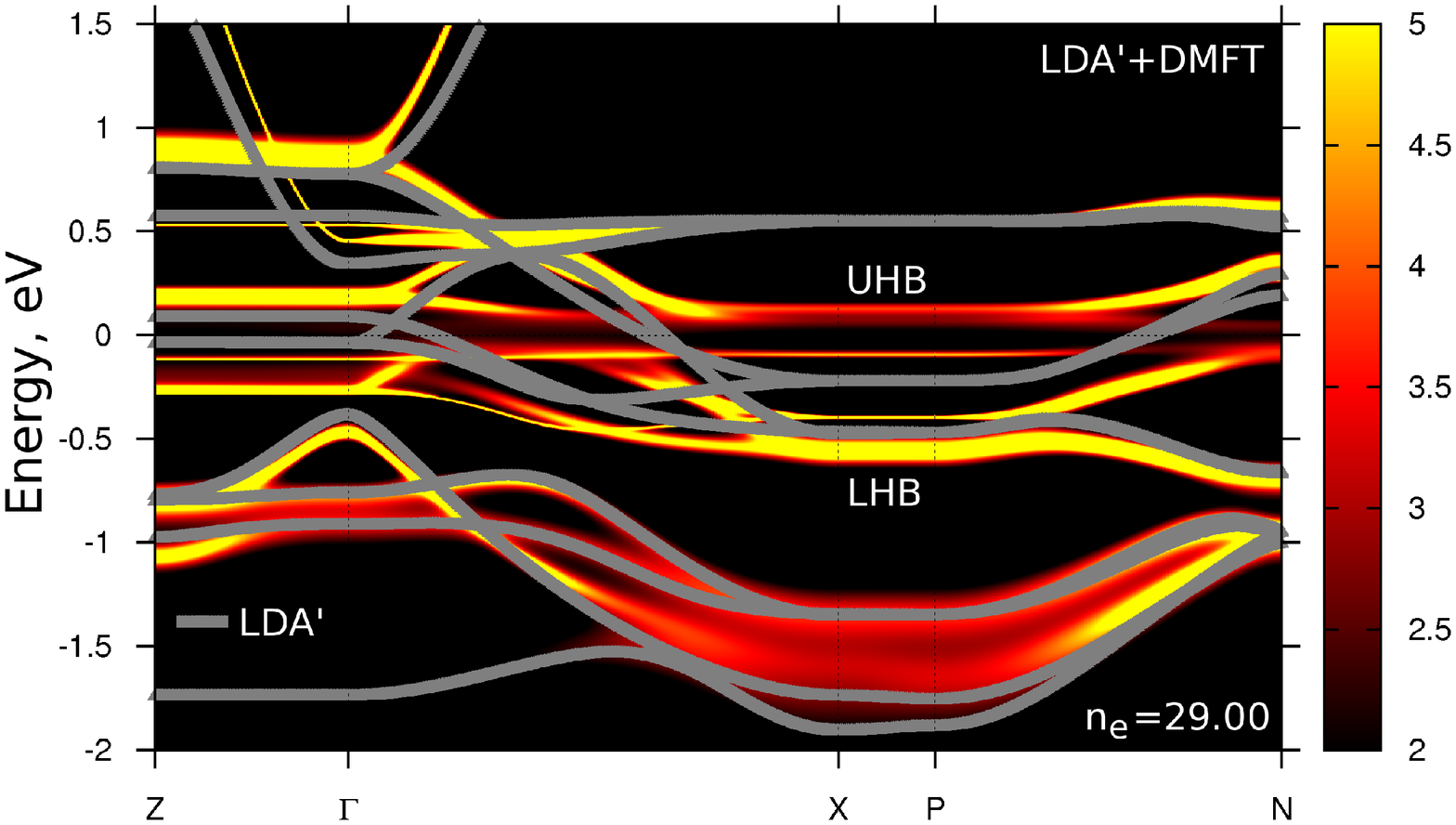}
\caption{Comparison of LDA$'$+DMFT calculated spectral function maps with renormalized by compression factor of 1.3 LDA$'$ bands (grey lines)
for K$_{1-x}$Fe$_{2-y}$Se$_2$ for different hole doping levels: n$_e$=29, 28, 27.2, 26.52 (from bottom to top)
along high symmetry direction of the first Brillouin zone. Fermi level is zero.}
\label{fig1}
\end{figure}

\section{Computational details}
\label{cd}

The K$_x$Fe$_2$Se$_2$ system is isostructural
to Ba122 pnictide (for the last one see Ref.~[\onlinecite{Nekr2}])
with ideal body centered tetragonal space group I4/mmm.
The K$_x$Fe$_2$Se$_2$ has $a$=3.9136\AA~ and $c$=14.0367\AA~
with K ions occupying $2a$, Fe -- $4d$ and Se -- $4e$ positions
with $z_{Se}$=0.3539 [\onlinecite{Guo10}].
This crystal structure was used in band structure calculations for
K$_{0.76}$Fe$_{1.72}$Se$_2$ within the linearized muffin-tin orbitals method
(LMTO)~[\onlinecite{LMTO}] using default settings [\onlinecite{Nekr_kfese}].

To take into account local Coulomb correlations we apply here
LDA$^\prime$+DMFT [\onlinecite{CLDA,CLDA_long}] approach which is
a modification of the well known LDA+DMFT method [\onlinecite{LDADMFT}].
LDA+DMFT Hamiltonian is usually written as
\begin{eqnarray}
\hat{H} &=& \hat{H}^{\rm LDA} + \hat{H}^{\rm Hub} - \hat{H}^{DC}.
\label{HLDADMFT}
\end{eqnarray}
The general problem with LDA+DMFT is that some portion of local electron-electron interaction
for presumed correlated $d$ shells is already included in the standard LDA ($\hat{H}^{\rm LDA}$).
To avoid its double counting due to Hubbard interaction
$\hat{H}^{\rm Hub}$ one has to subtract from $\hat{H}^{\rm LDA}$
the so called double counting correction term $\hat{H}^{DC}$.
(Explicit expressions for $\hat{H}^{\rm LDA}$ and $\hat{H}^{\rm Hub}$ can be found elsewhere [\onlinecite{CLDA_long}].)
The LDA$^\prime$+DMFT approach is the new attempt to solve the double counting problem, which is due to the
absence of universal expression for $\hat{H}^{DC}$, since there is no explicit microscopic (or diagrammatic) link
between the model (Hubbard like) Hamiltonian approach and the standard LDA.

To be short the main idea of the LDA$^\prime$+DMFT is {\em to exclude explicitly} from
the charge density the contribution of the presumably strongly correlated $d$ shells on the initial step of
LDA calculations. Then this redefined charge density (for some chosen orbital basis $\varphi_{i}({\bf r})$)
\begin{equation}
\rho^\prime ({\bf r})=\sum_{i\neq i_d}|\varphi _{i}({\bf r})|^{2}
\label{rhophi1}
\end{equation}
is used to calculate the local exchange-correlation energy $E_{\rm xc}^{\rm LDA}$ in LDA
and perform the self-consistent LDA band structure calculations for correlated bands at the
LDA stage of LDA+DMFT. Further on the local $d-d$ electron correlations are taken into account within the DMFT.
All states not counted as strongly correlated are then treated with the full power of DFT/LDA and
{\em full} $\rho$ in $E_{\rm xc}^{LDA}$.

Once this LDA$^\prime$ calculations with redefined charge density were done,
what is left for the correlated states out of interaction on the LDA stage would be
just the Hartree contribution which can be written in the fully localized limit form (FLL),
and which is the most consistent definition of double counting term here (other forms can be also used [\onlinecite{CLDA_long}]):
\begin{equation}
\hat{H}_{FLL}^{DC}=\frac{1}{2}U n_{d}(n_{d}-1)-
\frac{1}{2}{J}\sum_\sigma n_{d\sigma}(n_{d\sigma}-1),
\label{ELDAU}
\end{equation}
where $n_{d\sigma}=\sum_{m}n_{il_{d}m\sigma}=\sum_{m}\langle \hat{n}%
_{il_{d}m\sigma}\rangle $ is the total number of
electrons on strongly interacting orbitals and number of electrons per spin, $n_d=\sum_\sigma n_{d\sigma}$.

Effective five orbital impurity problem for K$_{1-x}$Fe$_{2-y}$Se$_2$  within DMFT was solved by Hirsh-Fye
Quantum Monte-Carlo algorithm [\onlinecite{QMC}],
at temperature 280K. LDA$^\prime$+DMFT
densities of states and spectral functions were obtained as discussed in
Ref.~[\onlinecite{CLDA}]. Coulomb parameters were taken to be $U$=3.75~eV and $J$=0.6~eV respectively [\onlinecite{exp}],
which are very close to calculated ones [\onlinecite{param}]. To define DMFT lattice problem we used the
full (i.e. without any downfolding or projecting) LDA Hamiltonian, which included all Fe-3d, Se-4p and K-4s states.

\section{Results and discussion}
\label{rad}

In Fig.~1 there we present the comparison of LDA$'$+DMFT calculated spectral function maps
in the wide energy window $\pm$2 eV along high symmetry directions in the first Brillouin zone with
the renormalized LDA$'$ bands (grey lines) for K$_{1-x}$Fe$_{2-y}$Se$_2$ at different hole doping levels n$_e$.
Renormalization (bandwidth compression) factor of LDA$'$ bands here is only 1.3, so that bandwidth renormalization due
to correlations is rather weak. The lower panel of Fig.~1 LDA$^\prime$+DMFT shows data for stoichiometric KFe$_{2}$Se$_2$
compound with total number of valence electrons n$_e$=29. For this composition all quasiparticle bands are rather well defined
for this wide energy range around the Fermi level. For hole doped cases with n$_e$=28, 27.2 and 26.52
(2nd, 3rd and 4th panels from bottom correspondingly) we see that Fe-3d bands obtained from LDA$^\prime$+DMFT become less pronounced
with hole doping. Overall rigid shift of LDA$'$ bands from stoichiometric case down to the most hole doped one is about 0.3 eV.

In Fig.~2 we show LDA$^\prime$+DMFT spectral function maps in the vicinity (-0.5 -- 0.2 eV) of the Fermi level with dominant
orbital character of quasiparticle bands denoted explicitly by black squares for $xz,yz$, black circles for $xy$,
white circles for $3z^2-r^2$ and white squares for $x^2-y^2$. We can see thet the orbital characters and forms of quasiparticle bands in this energy
interval (-0.1 -- 0.1 eV) change with the increase of hole doping, although characters of quasiparticle bands located outside this region
remain the same. Main orbital character of bands crossing the Fermi level is $xz,yz$ and $xy$. Also from Fig.~2 one can conclude that close enough to
the Fermi level (for all hole dopings) there are uniformly no well defined  quasiparticle bands (though some low intensity maxima of spectral density
can still be seen). This fact tells that the K$_{1-x}$Fe$_{2-y}$Se$_2$ for all hole dopings
is more correlated than 122 pnictide system.
\begin{table*}[!hb]
\caption{Quasiparticle energy scale renormalization factors and corresponding energy shifts (in eV, in round brackets) for
different bare Fe-3d LDA$'$ orbitals for all hole doping levels n$_e$
in the LDA$'$ scale energy interval from -1.0 eV to 0.4 eV.}
\begin{ruledtabular}
\begin{tabular}{|c|c|c|c|c|}
Orbital chracter  & n$_e$=26.52 &  n$_e$=27.20 & n$_e$=28.00 & n$_e$=29.00    \\
\hline
$xy$              & 1.5 (-0.23) &  3.9 (-0.73) & 2.65 (-0.61) & 1.7 (-0.35)   \\
\hline
$xz,yz$ (1)       & 4.2 (-0.78) &  3.0 (-0.75) & 2.6 (-0.69) & 1.7 (-0.38)   \\
\hline
$xz,yz$ (2)       & 2.3 (-0.48) & ~2.5 (-0.60) & 2.6 (-0.69) & 1.7 (-0.38)   \\
\hline
$xy,xz,yz$        & 1.2 (-0.10) &  1.3 (-0.10) & 1.3 (-0.10) & 1.4 (-0.17)   \\
\hline
$3z^2-r^2$        & 4.7 (-0.85) &  2.0 (-0.30) & 1.3 (-0.03) & 1.25 (0.0)   \\
\end{tabular}
\end{ruledtabular}
\label{tab1}
\end{table*}

This ``pseudogap'' like behaviour can be explicitly observed in Fig.~3 for all Fe-3d orbitals, where orbitally resolved
bare LDA$'$ and LDA$^\prime$+DMFT densities of states (DOS) for all four hole doping levels are presented.
Also by watching these DOSes one can actually see, that upon hole doping correlation effects become stronger.
This fact manifests itself in a different way for orbitals of various symmetry. First of all, for all Fe-3d orbitals we observe
narrowing of DOSes. For $3z^2-r^2$ (third panel from top) and $x^2-y^2$ (upper panel) this
narrowing is most evident at $\pm$1 eV and $\pm$0.5 correspondingly.
For $xz,yz$ (second panel from top) and $xy$ (bottom panel) increase of narrowing with doping is mostly concentrated
within the energy interval $\pm$0.4 eV.

To get more insight into the LDA$^\prime$+DMFT self-energy effects on bare LDA$^\prime$ bands we have
determined energy scale renormalizations and energy shifts for variety of separate dispersions of bare LDA$^\prime$ band structure
depicted in Fig.~4, which rather accurately fit bare bands to those plotted in the Fig.~2.
Also in Fig.~4 we show the standard LDA bands (dashed lines), just to emphasize that LDA$'$ dispersions are quite close to LDA ones
(see also Refs.~[\onlinecite{CLDA,CLDA_long,kfese_dmft}]).
The obtained energy scale renormalization (bandwidth compression) factors and energy shifts (shown in brackets) are collected in the Table~1
for all hole doping levels. These results show more complicated picture of bare LDA$'$ dispersion transformations, than those obtained in
Ref.~[\onlinecite{kfese_dmft}], where we have proposed that all Fe-3d band dispersions for the K$_{0.76}$Fe$_{1.72}$Se$_2$ should be just
compressed by a factor of 5 to get reasonable agreement with experiment. In contrast to our previous work
Ref.~[\onlinecite{kfese_dmft}] here we fit LDA$^\prime$ bare bands exactly to maxima positions of the spectral function
(see stars on Fig.~2). Thus we get more detailed picture which in general agrees with our early conclusions of
Ref.~[\onlinecite{kfese_dmft}].

Here we actually see, that different parts of bands in the first Brillouin zone acquire different (bandwidth or mass) renormalization factors,
which change with doping.
For example the $xy,xz,yz$  band (see Fig.~4) almost does not have doping dependence at all (see Table~1). Renormalization for the part of
$xz,yz$  band (2) only  slightly depend on doping. However, the $xz,yz$ band (1) becomes monotonically more correlated
(renormalization factor grows up to 4.2) while hole doping increase.
Correlation renormalization of the $xy$ orbital demonstrates nonmonotonic behaviour and is more pronounced for intermediate dopings.
The renormalization factor of the $3z^2-r^2$ band near the $\Gamma$ point abruptly increase up to 4.7 at n$_e$=26.52.
Here one should note that all four doping levels were treated with the same Coulomb interaction parameters.
Despite this fact upon doping several orbitals of K$_{1-x}$Fe$_{2-y}$Se$_2$ become more correlated (narrowed).
This phenomenon is apparently related to the change of the correlated orbitals occupancy.
The data collected in the Table~1 might be helpful for interpretation of APRPES spectra in simple terms of renormalized bare LDA band dispersions.

\section{Conclusions}
\label{concl}

In this paper we have performed detailed LDA$^\prime$+DMFT study of correlation effects in
K$_{1-x}$Fe$_{2-y}$Se$_2$ system at four hole doping levels: from
hypothetical stoichiometric composition with 29 valence electrons per unit cell
through intermediate values of valence electrons 28 and 27.2 towards composition
K$_{0.76}$Fe$_{1.72}$Se$_2$ with 26.52 electrons per unit cell, for which there are available
ARPES data on electronic dispersions [\onlinecite{exp}].

Within DMFT correlation effects are concentrated in the self-energy which
provides two types of modifications of the bare spectra at each energy: broadening (lifetime effects)
by imaginary part of self-energy and energy shift due to its real part.
We have shown, that in rather wide energy window $\pm$2 eV
(in terms of LDA energy scale) for all dopings the lifetime effects are relatively weak and
renormalization (compression) of quasiparticle bandwidths remains nearly the same and
of the order of $\sim$1.3.
However near the Fermi level self-energy effects become more pronounced and complicated.
In particular, the renormalization (bandwidth compression) factor grows from 1.3 for stoichiometric
composition upto nearly 5 for K$_{0.76}$Fe$_{1.72}$Se$_2$ (for $xz,yz$ and $3z^2-r^2$ bands).
Though all calculations were done with the same value of Coulomb (Hubbard) interaction,
this increase of renormalization upon doping tells us that correlations also increase with doping.
Also following recent tendency of experimental ARPES papers where energy shifts and
and renormalization factors are determined separately for different bare LDA bands [\onlinecite{ARPES_AFeSe,Kord}]
to fit ARPES data, we have provided quasiparticle mass renormalizations and energy shifts,
which rather accurately transform bare Fe-3d LDA$^\prime$ bands of various symmetries into
LDA$^\prime$+DMFT quasiparticle bands in different regions of the first Brillouin zone.
In fact, the detailed picture of band renormalizations close to the Fermi level in K$_{1-x}$Fe$_{2-y}$Se$_2$
is pretty complicated. For example, the $xy,xz,yz$ bands near the $\Gamma$-point and the $xz,yz$(2) bands in the middle of $\Gamma$-X and P-N directions
almost do not feel doping changes. On the opposite, the $xz,yz$(1) bands in all directions and
the $3z^2-r^2$ band near the $\Gamma$-point become monotonically more
correlated (renormalized) with doping increase. Finally, correlation renormalization of the $xy$ band demonstrates
non monotonic behaviour with doping, becoming  more correlated for intermediate dopings.
However, the general conclusion of Ref. [\onlinecite{Nekr_kfese}] remains valid K$_{1-x}$Fe$_{2-y}$Se$_2$ systems demonstrates
more pronounced correlation effects in contrast to 122 iron pnictides. This is clearly reflected in the absence of well defined
quasiparticle bands in the vicinity of the Fermi level, which demonstrates a kind of ``pseudogap'' behavior (``dark'' region around
the Fermi level in Fig. 2). The interesting problem for the future studies is possible manifestation of these effects in optical
conductivity, as well as their role in the formation of superconducting state.

\section{Acknowledgements}

We thank A.I. Poteryaev for providing us QMC code and many helpful discussions.
This work is partly supported by RFBR grant 11-02-00147 and was performed
within the framework of programs of fundamental research of the Russian
Academy of Sciences (RAS) ``Quantum mesoscopic and disordered structures''
(12-$\Pi$-2-1002) and of the Physics Division of RAS  ``Strongly correlated
electrons in solids and structures'' (012-T-2-1001).
NSP acknowledges the support of the Dynasty Foundation and International Center
of Fundamental Physics in Moscow. DMFT(QMC) calculations were performed on
supercomputer ``Uran'' in the Institute of Mathematics and Mechanics of UB RAS.

\begin{figure}
\includegraphics[width=.5\textwidth]{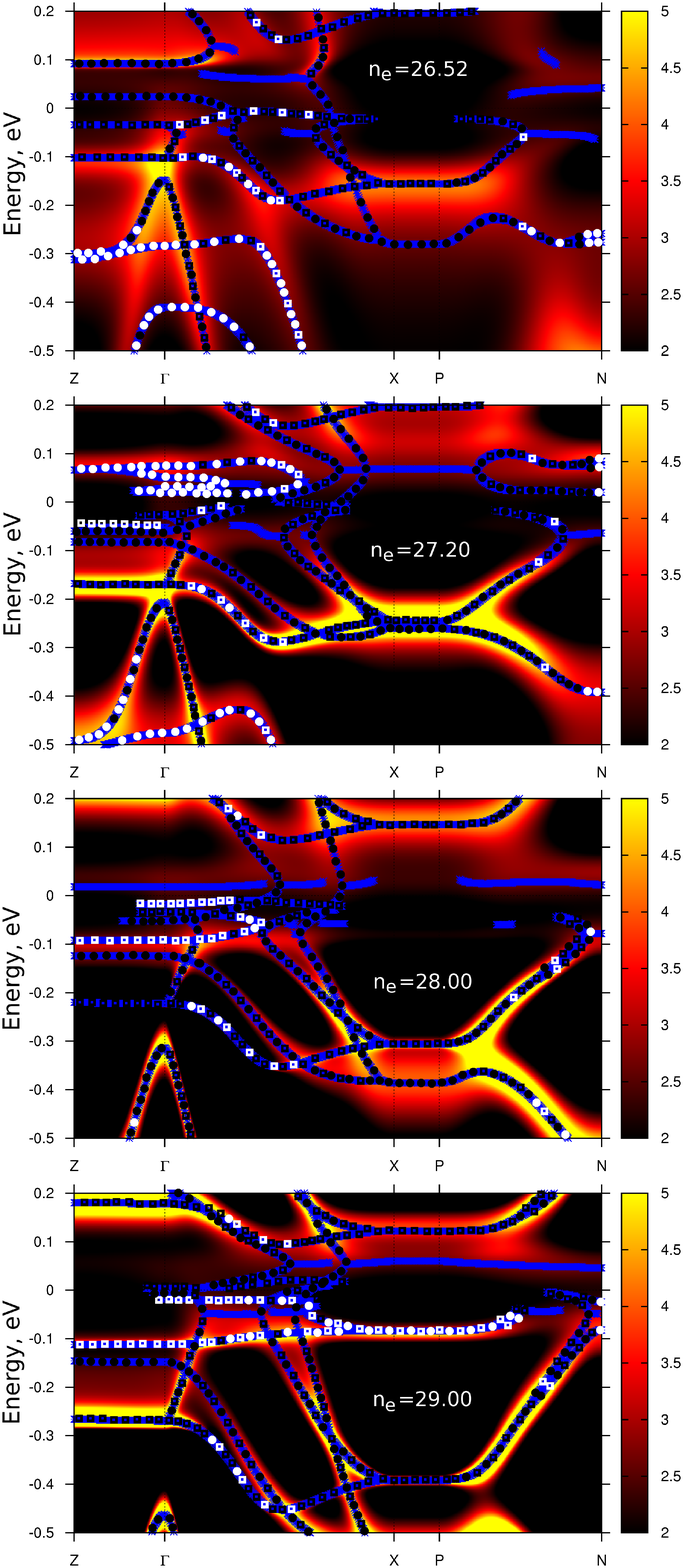}
\caption{LDA$'$+DMFT spectral density maps along high symmetry directions of the first Brillouin zone.
Maxima of intensity are shown by stars.
Dominant orbital character of quasiparticle bands is marked as follows: black squares - $xz,yz$, black circles - $xy$,
white circles - $3z^2-r^2$, white squares - $x^2-y^2$. Fermi level is zero.}
\label{fig2}
\end{figure}

\begin{figure}
\includegraphics[width=.5\textwidth]{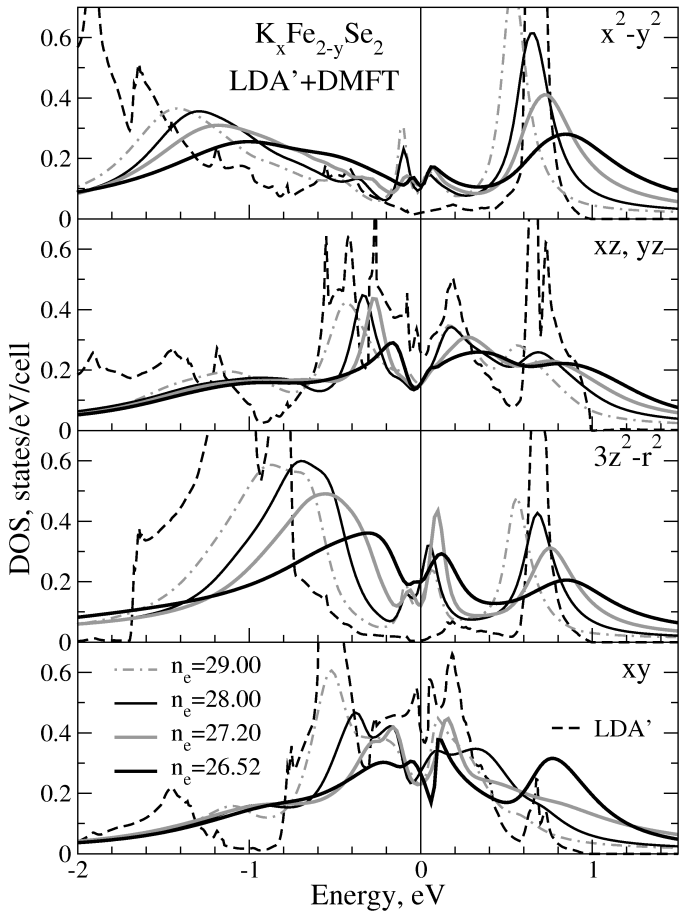}
\caption{Comparison of LDA$'$ (dashed black line) and LDA$'$+DMFT density of states (DOS) for
K$_{1-x}$Fe$_{2-y}$Se$_2$  for different dopings n$_e$.
Thick black line corresponds to n$_e$=26.52, thick grey line -- n$_e$=27.20, thin black line -- n$_e$=28.00, grey dot-dashed line -- n$_e$=29.00.
Fermi level is zero.}
\label{fig3}
\end{figure}

\begin{figure}
\includegraphics[width=.5\textwidth]{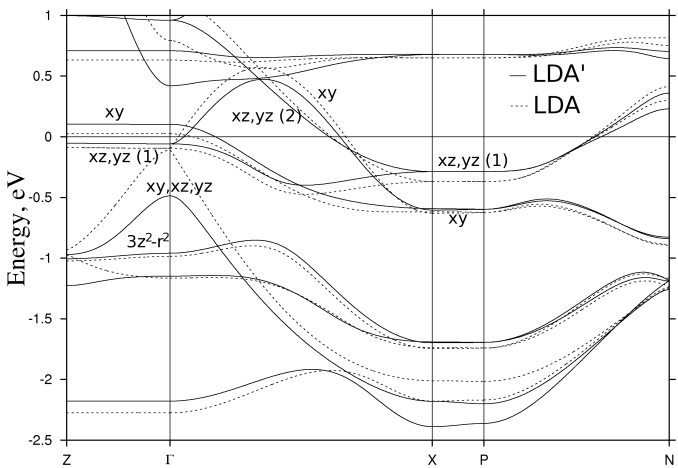}
\caption{Bare LDA$'$ (full line) and LDA (dashed line) band dispersions for stoichiometric KFe$_2$Se$_2$ system
with orbital characters explicitly shown. Numbers in brackets after orbital symmetry symbol
enumerate corresponding parts of the bands in the first Brillouin zone (see Table~1).}
\label{fig4}
\end{figure}

\end{document}